\documentclass{nature}

\usepackage{amsmath}
\usepackage[pdftex]{graphicx}
\usepackage{dcolumn}  
\usepackage{bm}          
\usepackage{dsfont}
\usepackage{color}
\usepackage{braket}
\usepackage{amssymb}
\usepackage{booktabs}
\usepackage[colorlinks, 
linkcolor=black,
anchorcolor=black,
citecolor=green,
urlcolor=black
]{hyperref}

\newcounter{defcounter}
\usepackage{amsthm}

\title{Conical Intersections Enable Ultrafast Molecular Spin Control in a Chromium Complex } 

\author{Zihui Liu$^{1,*}$, Junhua Zhou$^{1,*}$, Tianrui Chen$^{1}$, Michael Penny$^{2}$, Sara Mosca$^{3}$, Mengyuan Cui$^{1}$, Vandana Tiwari$^{4}$, R. J. Dwayne Miller$^{5}$, Fulu Zheng$^{1}$, Ajay Jha$^{2,6,7}$, Hong-Guang Duan$^{1}$} 

\begin{document} 

\maketitle 

\begin{affiliations} 
\item Department of Physics, School of Physical Science and Technology, Ningbo University, Ningbo, 315211, P.R. China 
\item Rosalind Franklin Institute, Harwell, Oxfordshire OX11 0QX, United Kingdom
\item Central Laser Facility, Research Complex at Harwell, STFC Rutherford Appleton Laboratory, UK Research and Innovation (UKRI), Harwell, Oxford OX11 0QX, United Kingdom
\item Linac Coherent Light Source, SLAC National Accelerator Laboratory, Menlo Park, California 94025
\item The Departments of Chemistry and Physics, University of Toronto, 80 St. George Street, Toronto M5S 3H6, Canada
\item Department of Pharmacology, University of Oxford, Oxford, OX1 3QT United Kingdom
\item Research Complex at Harwell, Rutherford Appleton Laboratory, Didcot OX11 0QX, United Kingdom
  \\ 
\centerline{\underline{\date{\bf \today}}} 
\end{affiliations} 

\begin{abstract} 

Molecular spintronics seeks to control spin states in single molecules for ultrafast switching and efficient information processing. Transition metal complexes are promising candidates for such applications due to their modular ligand fields, diverse spin configurations, and potential for spin-vibronic coupling that facilitates rapid spin dynamics. Chromium(III) complexes, in particular, offer long-lived emissive doublet states and chemical robustness, making them attractive for room-temperature spin control. Here we investigate the spin-state dynamics of tris(2,4-pentanedionato)chromium(III), [Cr(acac)$_{3}$], a photochemically stable d$^{3}$ complex with minimal vibrational congestion. Using ultrafast transient grating and two-dimensional electronic spectroscopy with $\sim$10 fs resolution, we directly probe vibrational and electronic dynamics associated with the $^{4}$T$_{2}$ $\rightarrow$ $^{2}$E intersystem crossing (ISC). These measurements reveal coherent vibrational modes implicated in mediating nonadiabatic spin transitions. Complementary theoretical modelling shows that vibronic coupling and spin-orbit interactions promote the formation of multiple conical intersections, providing ultrafast channels for spin-flip dynamics. Metal-ligand bending and stretching modes serve as tuning and coupling coordinates, enabling ISC despite weak spin-orbit coupling in 3d transition metal. Our study provides mechanistic insight into spin-vibronic dynamics in Cr(III) complexes and establishes a design framework for achieving ultrafast molecular spin switching, advancing the development of optically addressable spin centres for future spintronic and quantum technologies.

\end{abstract} 


\section*{Introduction} 
Molecular spintronics seeks to harness the quantum properties of electron spin in single molecules for next-generation information technologies that are ultrafast, energy-efficient, and miniaturised to the molecular scale.\cite{Ref1,Ref2} At the core of this field lies the challenge of achieving coherent spin manipulation and state interconversion compatible with device operation, at ever increasing demands on speed.\cite{Ref3,Ref4} Optical excitation offers a non-invasive and ultrafast route to perturb and read out spin states, enabling light-induced spin switching in paramagnetic systems. Recent experiments have demonstrated all-optical spin control in molecular systems on sub-nanosecond timescales, providing compelling proof-of-concept for spintronic switching at the molecular level.\cite{Ref5} These advances have propelled transition metal complexes, particularly those based on first-row elements, into the spotlight as adaptable spin carriers.\cite{Ref6,Ref7,Ref8,Ref9,Ref10,Ref11,Ref12,Ref13,Ref14,Ref15,Ref16,Ref16a} Their intrinsic electronic degeneracy, variable oxidation and associated spin states, and stong ligand-field interactions allow them to host a wide range of spin manifolds. These factors can be readily modulated by chemical structure and external fields, making them particularly attractive for the design of spin-functional molecular materials.\cite{Ref16b}

Within this broad class, Chromium(III) complexes have emerged as especially promising candidates. As kinetically inert, octahedrally coordinated species with a d$^3$ configuration, they exhibit exceptional chemical robustness, which is a prerequisite for stable device integration.\cite{Ref16} The prototypical “molecular ruby” systems, such as [Cr(Al$_2$O$_3$)], are well known for their sharp ligand-field transitions and long-lived spin-flip excited states, which persist from hundreds of microseconds to milliseconds in favourable host environments.\cite{Ref17} The $^2$E doublet state, accessed via photoexcitation from the quartet $^4$A$_2$ ground state, possesses minimal orbital angular momentum, which suppresses nonradiative decay and enhances spin coherence lifetimes.\cite{Ref18} These features allow Cr(III) complexes to retain spin polarisation long after photoexcitation, making them viable for room-temperature optical spin storage and manipulation. Moreover, their modular coordination chemistry enables systematic tuning of ligand-field strengths and geometries, offering a designable platform for exploring spintronic function. Thus, understanding the spin dynamics of Cr(III) complexes is not only of fundamental interest, but also offer an important platform for the rational design of optically triggered molecular spin qubits.

Central to the development of such spin-functional molecules is the mechanistic elucidation of ISC processes, which are nonadiabatic transitions between states of different spin multiplicity.\cite{Ref19,Ref20,Ref21,Ref22,Ref23,Ref24,Ref25,Ref26,Ref27,Ref27a,Ref27b,Ref27c,Ref27d,Ref27e,Ref27f} In Cr(III) complexes, the key ISC event is the photoinduced $^4$T$_2$ $\rightarrow$ $^2$E transition, which converts an optically excited quartet state into a metastable doublet.\cite{Ref18,Ref28} While spin-orbit coupling (SOC) is inherently weak in 3d metals, the presence of vibronic interactions, particularly via ligand-metal bond distortions, can enhance ISC rates dramatically.\cite{Ref27,Ref29,Ref29a} In tris(2,4-pentanedionato)chromium(III), Cr(acac)$_3$, ultrafast spectroscopic studies have revealed persistent coherent oscillations at 164 cm$^{-1}$, corresponding to a Cr–O stretching mode that has been proposed to couple the quartet and doublet surfaces.\cite{Ref30} This observation substantiates a vibronically mediated ISC mechanism, wherein nuclear motion actively guides the system through regions of strong spin-state mixing on the potential energy surface. Building on these observations, a key unresolved challenge is to establish a unified, mechanistic picture linking vibronic coupling, SOC, and potential energy surface topology in governing spin conversion, an interplay that remains insufficiently resolved. Achieving this is crucial not only to rationalize ultrafast spin dynamics, but to enable predictive control of spin switching in 3d transition metal complexes for functional spintronic applications.

To investigate ultrafast spin-switching dynamics in detail, we employ [Cr(acac)$_3$], as a model system. This complex combines photochemical inertness with minimal vibrational congestion, owing to the pseudoaromaticity of its planar chelate rings and the compact nature of its acetylacetonate ligands.\cite{Ref31,Ref32} These features make it particularly amenable to time-resolved spectroscopic analysis, facilitating the clear assignment of vibrational coherences. Moreover, the excited-state dynamics of [Cr(acac)$_3$] have been well characterised in prior work using transient absorption spectroscopic studies, providing a firm foundation for our present focus on spin-vibronic interplay.\cite{Ref18,Ref30,Ref33,Ref34,Ref35,Ref36} In this study, we explore the mechanisms governing intersystem crossing in [Cr(acac)$_3$] using ultrafast coherent spectroscopies: transient grating (TG)\cite{Ref37,Ref38} and two-dimensional electronic spectroscopy (2DES)\cite{Ref39,Ref40,Ref41,Ref42} with $\sim$10 fs excitation, as shown in Fig.\ \ref{fig:Fig1}(A). These techniques offer complementary sensitivity to vibrational coherence and electronic dynamics, enabling us to probe the coupling between nuclear and electronic degrees of freedom in real time.\cite{Ref43,Ref44,Ref45} By combining experiment with computational modelling of spin-orbit and vibronic interactions, we have characterized how vibrational modes mediate photoinduced ISC from $^4$T$_2$ $\rightarrow$ $^2$E (Fig.\ \ref{fig:Fig1}(B)). Our approach offers a multidimensional view of nonadiabatic spin dynamics in Cr(III), contributing to a deeper mechanistic understanding of ultrafast spin-state control in molecular systems.


\section*{Results and Discussions} 

\begin{figure}[h!]
\begin{center}
\includegraphics[width=15.0cm]{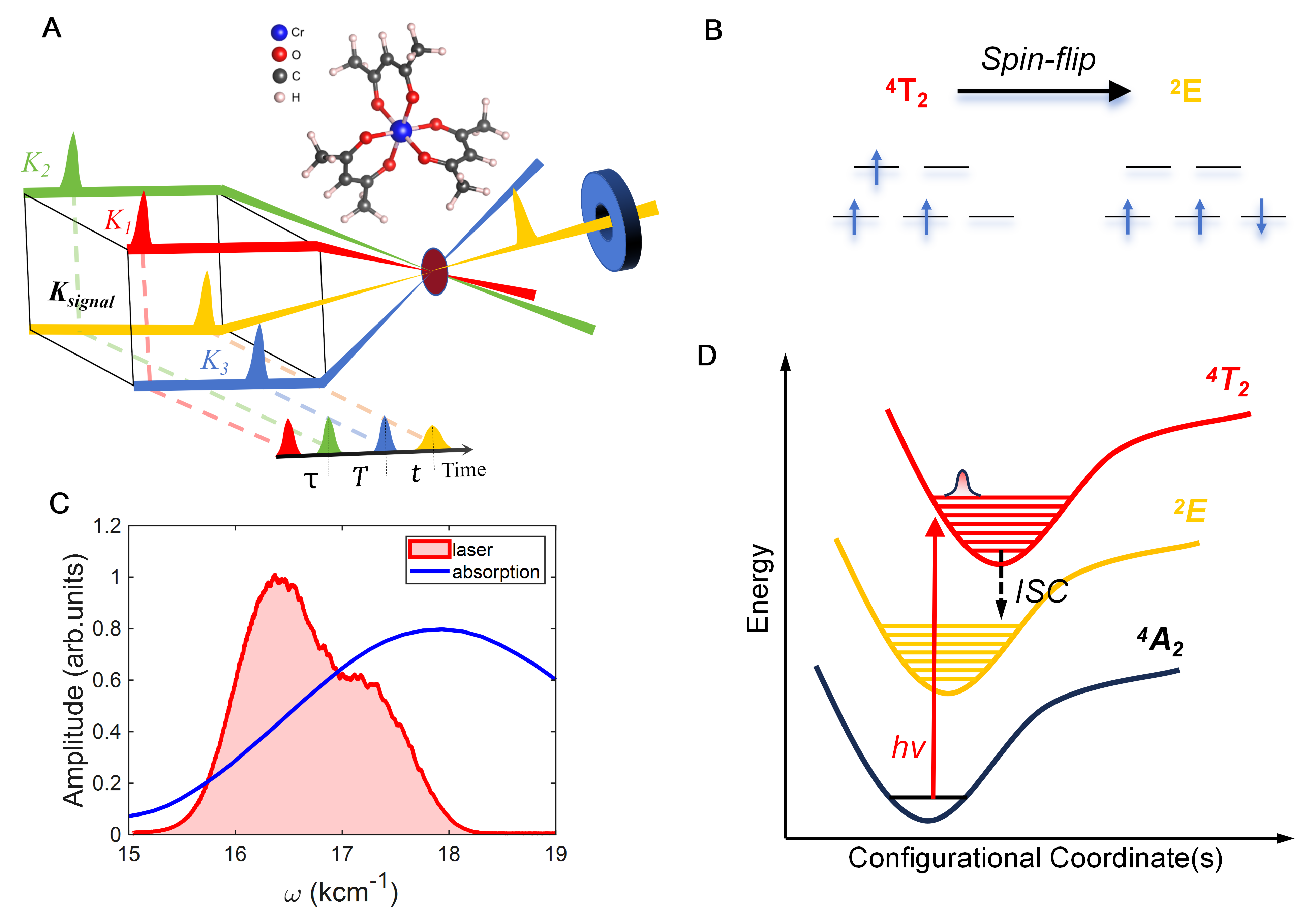}
\caption{\label{fig:Fig1} Photoinduced spin-state dynamics in [Cr(acac)$_{3}$]. (A) Schematic of the two-dimensional electronic spectroscopy (2DES) experiment used to probe spin dynamics in [Cr(acac)$_{3}$]. Three phase-controlled laser pulses (k$_{1}$–k$_{3}$) generate a third-order nonlinear signal, which is heterodyne-detected to resolve ultrafast electronic and vibrational coherences. The molecular structure of  [Cr(acac)$_{3}$] is shown. (B) Electronic configuration change associated with the spin-flip transition from the photoexcited quartet state ($^4$T$_2$) to the doublet state ($^2$E). (C) Steady-state absorption spectrum (blue) and excitation laser profile (red shaded), indicating spectral overlap used to initiate the transition. (D) Schematic potential energy surfaces of the ground ($^4$A$_2$), excited quartet ($^4$T$_2$), and doublet ($^2$E) states, showing photoexcitation and subsequent intersystem crossing.} 
\end{center}
\end{figure}

The experimentally recorded absorption spectrum of the complex, depicted as a solid blue line in Fig.\ \ref{fig:Fig1}(C), exhibits a prominent feature corresponding to the spin-allowed electronic transition (depicted in Fig.\ \ref{fig:Fig1}(D)). Superimposed on this plot is the laser excitation profile, represented by the red-shaded area, indicating spectral overlap sufficient to selectively initiate this electronic transition. To follow the excited state population dynamics, the relevant electronic energy levels of [Cr(acac)$_{3}$] are schematically illustrated in Fig.\ \ref{fig:Fig1}(D). Broadband TA measurements (Supplementary Information, Fig. S1) reproduce the excited-state signatures reported by McCusker and co-workers for [Cr(acac)$_3$]\cite{Ref30}, confirming the expected transient features. However, the ~200-fs time resolution and substantial spectral overlap, and spectral congestion due to inhomogenous broadening, limited the information needed for unambiguous state assignment. These constraints motivate shorter excitation pulses and frequency-resolved 2DES to disentangle overlapping states.

\subsection{Two-dimensional electronic spectroscopy} 

\begin{figure}[h!]
\begin{center}
\includegraphics[width=15.0cm]{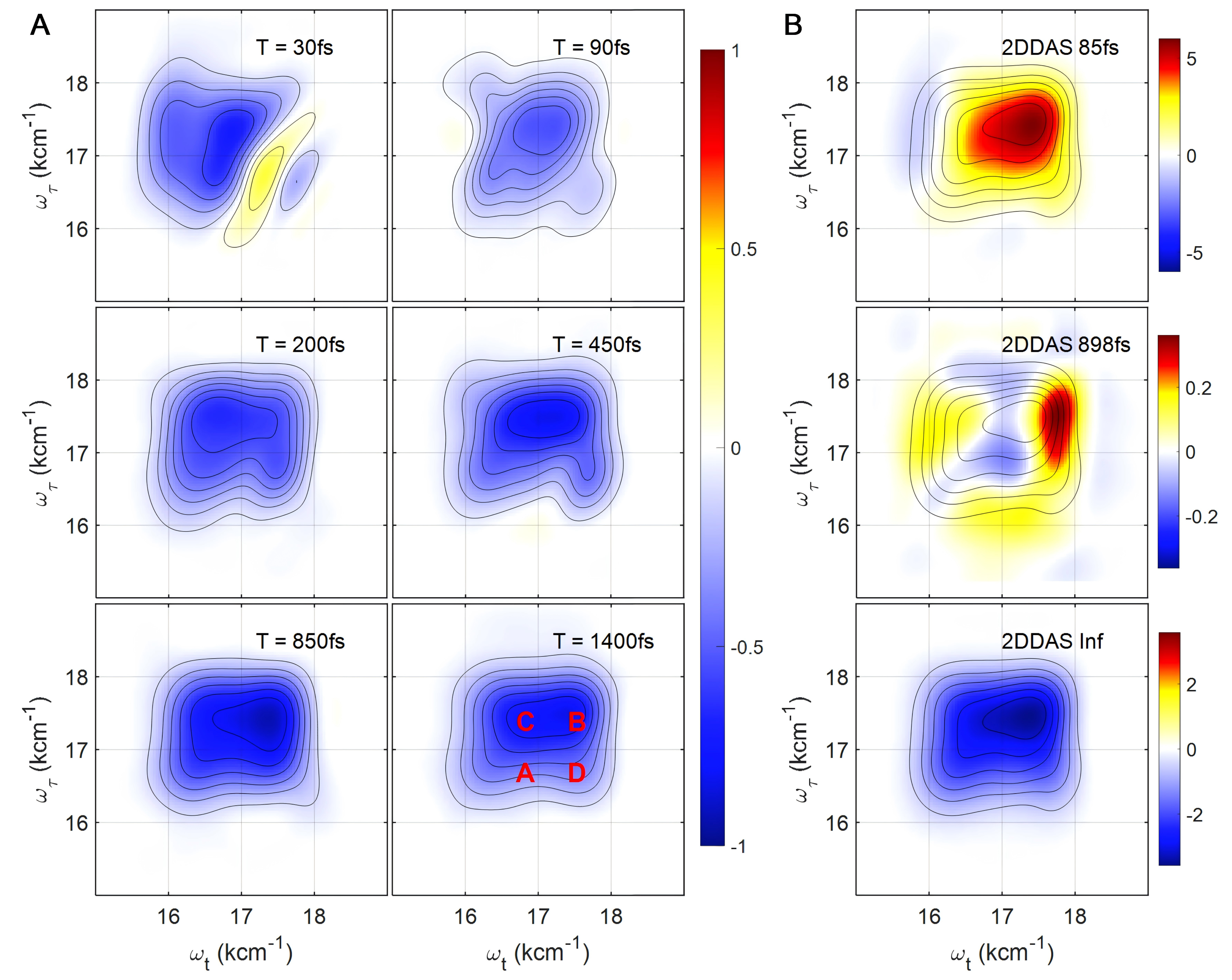}
\caption{\label{fig:Fig2} Two-dimensional electronic spectroscopy reveals ultrafast excited-state dynamics in [Cr(acac)$_{3}$]. (a) 2DES spectra recorded at selected waiting times (T = 30, 90, 200, 450, 850, and 1400 fs), showing the evolution of ground-state bleach (GSB, in red) and excited-state absorption (ESA, in blue) features. The spectra evolve from an initially congested lineshape at early times to a more structured and stabilized pattern at longer delays. Representative spectral positions (A–D) used for kinetic analysis are indicated at T = 1400 fs. (B) Two-dimensional decay-associated spectra (2DDAS) obtained from global analysis of the 2DES dataset, corresponding to characteristic lifetimes of 85 fs, 898 fs, and a long-lived (effectively infinite) component. } 
\end{center}
\end{figure}

We performed 2DES measurements to investigate the ultrafast excited-state dynamics of [Cr(acac)$_{3}$], with results summarized in Fig.\ \ref{fig:Fig2}. The 2D spectra were collected at different waiting times, enabling the temporal evolution of the excited-state features. At T=30 fs, the spectra carries details of the electronic dephasing between the ground and excited states. To minimize the pulse overlap effects, present at time-zero (T = 0 fs), we selected the 30 fs spectrum as the earliest reliable data point for this analysis. At this delay, we performed an anti-diagonal cut through the 2D spectral map to resolve the spectral line shape and extract the lifetime of electronic dephasing.\cite{Ref43} The resulting line shape was analyzed using a Lorentzian fitting model, from which we determined an electronic dephasing time of $\sim$16 fs, as shown in SI, Fig. S4 (also, SI, Section-II).

At T = 30 fs, 2DES data is characterized by a strong negative feature (excited state absorption, ESA) and an overlapping positive signal (ground-state bleach, GSB and stimulated emission, SE). At this short delay, the spectral overlap between GSB and ESA is significant, making it challenging to resolve underlying spectral features. As the waiting time increases to 90 fs, a distinct negative peak appears at coordinates ($\omega_{\tau}$, $\omega_{t}$) = (17500, 17000) cm$^{-1}$, flanked by two weaker negative sidebands, indicating the evolution of excited-state features. By T = 200 fs, the 2D spectrum evolves into a more complex pattern, with two principal peaks and two off-diagonal cross peaks. The negative peaks become increasingly pronounced beyond T = 450 fs, indicating further stabilization of the excited-state manifold. Notably, the spectra at 850 fs and 1400 fs show consistent spectral features, suggesting that the system reaches a relatively stable excited-state configuration by this time. The amplitude evolution of four representative peaks, labeled A, B, C, and D in 2DES data for T = 1400 fs are presented in SI in Fig S3. 

To further elucidate the dynamic processes underlying the 2D electronic spectra, we employed a global fitting analysis to extract decay lifetimes across the full 3D dataset of the 2DES, resolved along the waiting time T (details are presented in the SI, section-II(b)). The resulting 2D Decay Associated Spectra (DAS) components are presented in Fig.\ \ref{fig:Fig2}(B), corresponding to decay lifetimes of 85 fs, 898 fs, and a long-lived (effectively infinite) component. For reference, we superimpose contour lines of the 2DES at T = 600 fs to assist in correlating these features with the original spectra. The 85 fs component shows a strong positive peak alongside a smaller, negative side peak. This component represents the initial, fast relaxation dynamics of the excited state population. The high amplitude of this component underscores its significance in the early-time evolution of the system. The second DAS component is associated with a slower decay process characterized by a lifetime of 898 fs. This component displays a prominent negative peak at ($\omega_{\tau}$, $\omega_{t}$) = (17500, 17800) cm$^{-1}$, along with additional side peaks of lower intensity. It reflects a more gradual decay in this spectral region, indicating a secondary relaxation channel or vibrational redistribution process. Finally, the third component, corresponding to an infinite lifetime, captures the long-lived spectral features that persist beyond the measured time window. This component shares a similar spectral profile with the 2DES at extended waiting times, affirming the presence of a stable, equilibrated excited-state population.

\subsection{Coherent dynamics} 

\begin{figure}[h!]
\begin{center}
\includegraphics[width=13cm]{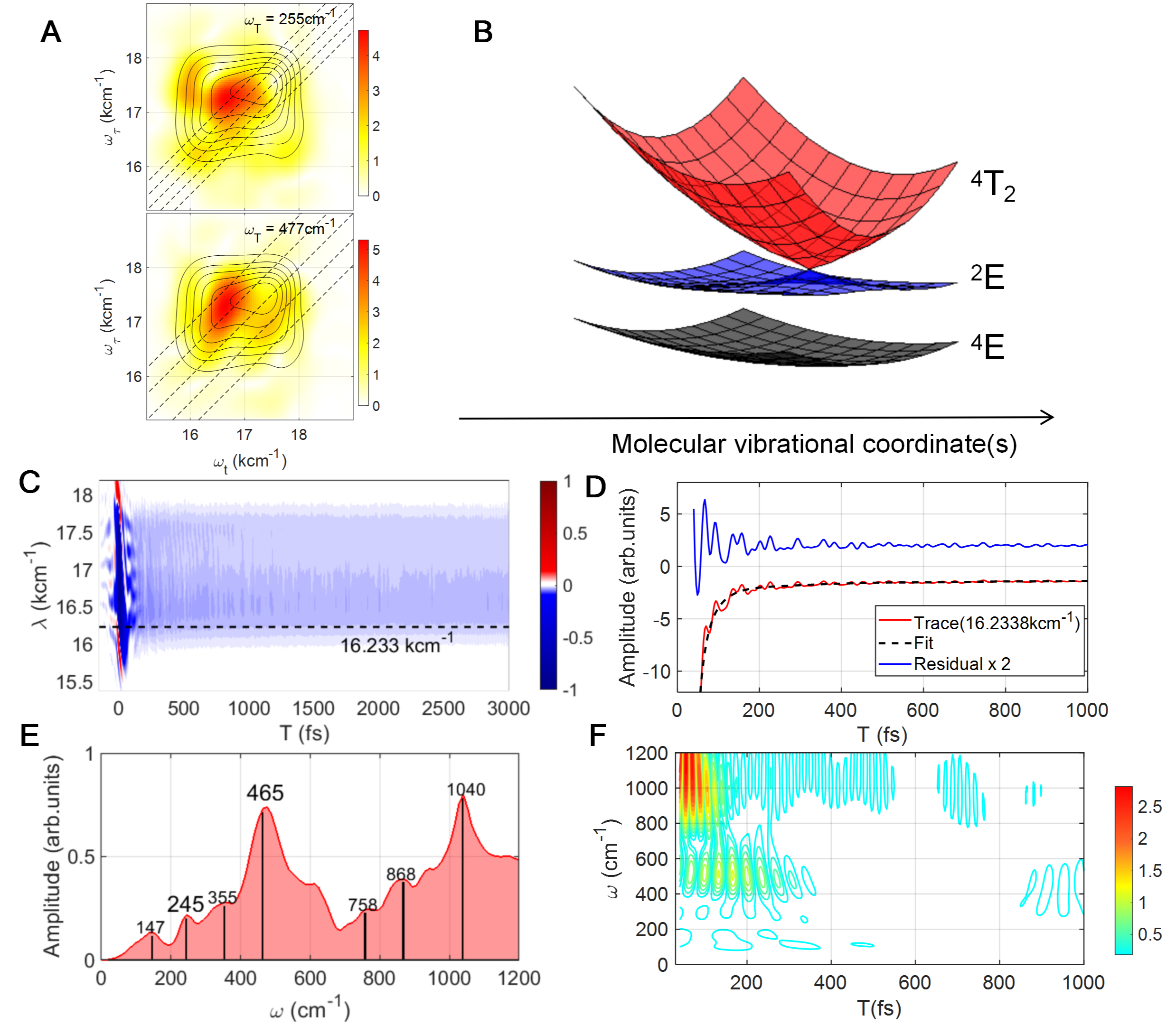}
\caption{\label{fig:Fig3} Vibronic coherences and time-frequency-resolved dynamics in [Cr(acac)$_{3}$]. (A) 2D vibrational maps (255, 477 cm$^{-1}$)  from Fourier-transformed 2DES residuals showing coherent mode distributions and electronic coupling. (B) Model potential energy surfaces illustrating nuclear-coordinate mediation of electronic states and active vibrational modes. (C) Transient grating spectrogram versus probe wavelength and waiting time; dashed line marks the selected probe. (D) Time trace at the chosen wavelength with experimental data, multi-exponential fit, and oscillatory residual isolating coherent dynamics. (E) Fourier spectrum of residuals identifying low- and high-frequency vibrational modes. (F) Wavelet analysis revealing the time–frequency evolution of vibronic coherences.} 
\end{center}
\end{figure}

To gain insight into the coherent vibrational dynamics underlying the electronic transitions, we extracted the residuals from the globally fitted 3D 2DES dataset and performed a Fourier transform along the waiting time axis T. This analysis yielded frequency-resolved 2D vibrational maps. The resulting maps, presented in Fig.\ \ref{fig:Fig3}(A), correspond to selected vibrational frequencies of 255 and 477 cm$^{-1}$ (additional frequency maps for 85, 187, 885, and 1242 cm$^{-1}$ are presented in SI, Figure S6). For reference, contour lines of the 2DES at T = 600 fs are overlaid to aid in correlating vibrational features with the electronic spectra. The vibrational map at 85 cm$^{-1}$ reveals a prominent central peak at ($\omega_{\tau}$, $\omega_{t}$) = (17100, 16600) cm$^{-1}$, indicating strong vibrational coherence near the absorption maximum. Similarly, the 187 cm$^{-1}$ map shows a central peak with smaller sidebands, highlighting vibrational progression, which is emphasized with dashed reference lines. At 255 cm$^{-1}$, a shifted peak is observed at (17200, 16700) cm$^{-1}$, while the 477 cm$^{-1}$ mode shows a strong off-center peak. Higher-frequency modes at 885 and 1242 cm$^{-1}$ display structured main and cross peaks aligned with vibrational transitions. These findings provide a detailed map of vibronic coherences, which might reflect evolution of molecular vibrational coordinates during the ISC process (Fig.\ \ref{fig:Fig3}(B)). 

To complement the 2DES analysis, we employ TG spectroscopy, which selectively enhances weak coherent oscillations, enabling clearer identification of vibrational dynamics. Fig.\ \ref{fig:Fig3}(C) displays the measured TG spectrum of [Cr(acac)$_{3}$], highlighting a pronounced ESA band. For quantitative analysis, we extracted the time-domain trace at $\omega$ = 16233 cm$^{-1}$, shown as a blue dashed line in Fig.\ \ref{fig:Fig3}(D), and conducted a Fourier transform, with the resulting frequency-resolved data plotted in Fig.\ \ref{fig:Fig3}(E). Oscillatory features were resolved at 147, 245, 355, 465, 753, 868, and 1040 cm$^{-1}$, marked by black lines for reference. The close correspondence between these frequencies and those observed in the Raman spectra (provided in the SI, Fig. S13) confirms the vibrational origin of these coherences and validates the accuracy of the TG measurements. Similar analysis have been performed for trace at $\omega$ = 17361 cm$^{-1}$ and shown in SI, Fig. S2. It revealed coherent oscillations, persisting for $\sim$1 ps and the vibrational modes resolved were found at 147, 257, 477, 685, 881, 1015, and 1150 cm$^{-1}$. These modes represent a combination of low-frequency metal-ligand stretching motions and higher-frequency ligand-centered vibrations, which may play important roles in mediating excited-state interactions.

To further investigate the temporal evolution and lifetimes of these vibrational modes, we applied wavelet transform analysis to the residual data (details in SI, section-II E). The results for trace at $\omega$ = 16233 cm$^{-1}$ is presented in Fig.\ \ref{fig:Fig3}(F). In The wavelet-transformed data reveals three prominent high-frequency modes centered around 1002, 1102, 1225 and 1378 cm$^{-1}$. These modes exhibit strong initial intensities but rapidly decay within 200 fs, indicating they are short-lived coherences generated impulsively by the excitation pulse. This fast decay may reflect energy dissipation through intramolecular vibrational redistribution (IVR) or coupling to other modes within the vibrational manifold. In contrast, a mid-frequency mode centered at 479 cm$^{-1}$ shows a distinct temporal profile. This mode emerges shortly after excitation, reaches maximum amplitude at T = 200 fs, and decays over the subsequent 200 fs. The delayed peak suggests it may result from coherent vibrational wavepacket formation driven by mode-selective coupling between the ground and excited states\cite{Mitra 2026}. Additionally, low-frequency modes in the 100-200 cm$^{-1}$ region persist for over 400 fs, reflecting their role as collective skeletal motions or metal-ligand deformations that are more robust against decoherence mechanisms. To extract the quantitative lifetimes of the observed vibrational modes, we fitted the residual oscillatory components using exponentially damped sine functions. These fits were performed using MATLAB 2022(b)’s curve fitting toolbox, and the initial frequency estimates were guided by Linear Predictive Singular Value Decomposition (LPSVD) \cite{lpsvd} analysis, which effectively deconvolves closely spaced frequencies. The full fitting procedure is elaborated in the SI (section II, D). The resulting lifetimes and corresponding frequencies of each vibrational mode are provided in Fig.\ S7-S10 in the SI. The associated calculations present the detailed motions of molecule, which will be described in the next section.

\subsection{Theoretical calculations} 

\begin{figure}[h!]
\begin{center}
\includegraphics[width=15.0cm]{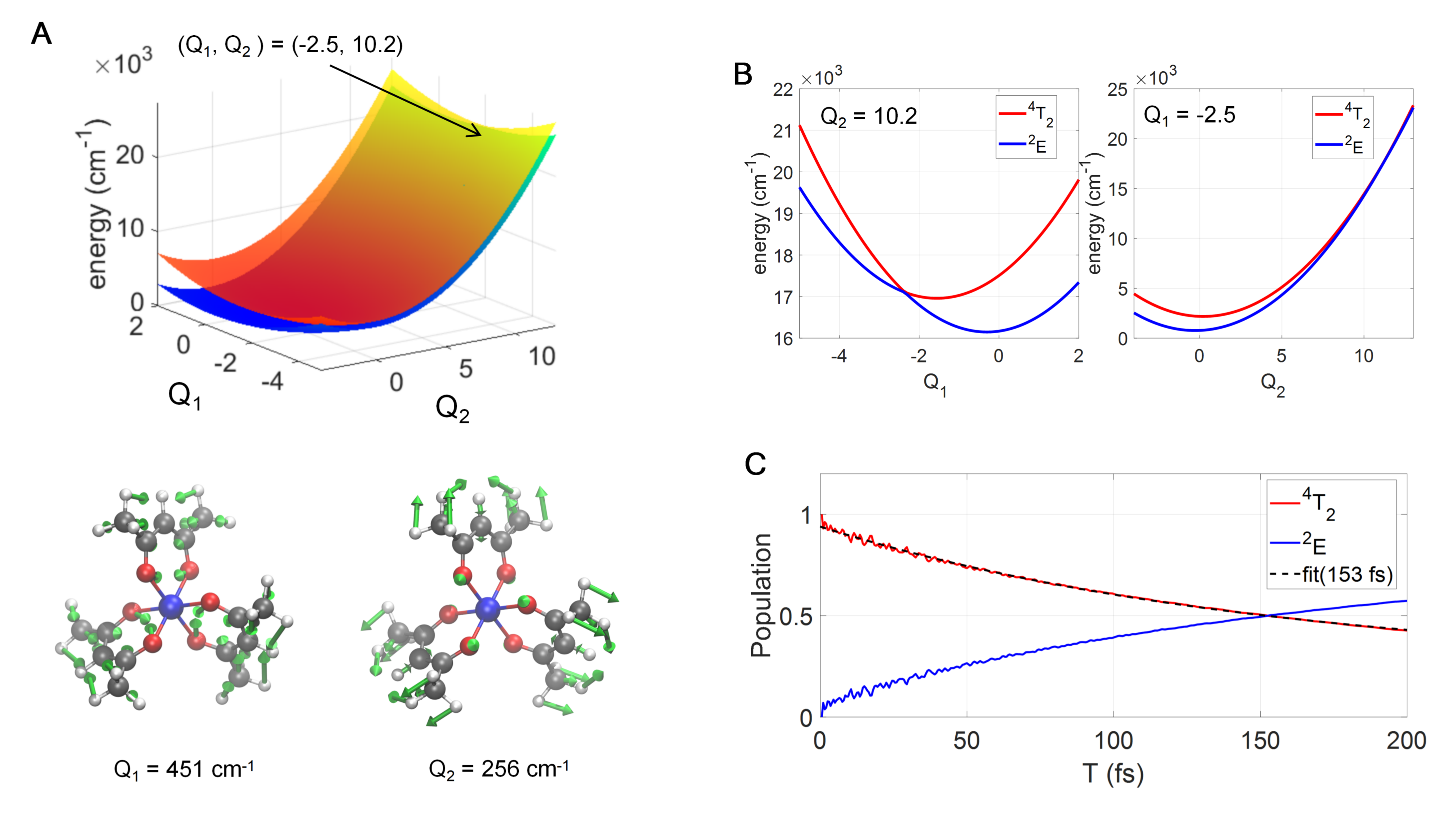} 
\caption{\label{fig:Fig4} Conical intersection mediated by vibrational coordinates in [Cr(acac)$_{3}$]. (A) Calculated potential energy surfaces of the $^4$T$_2$ and $^2$E states along two vibrational coordinates (Q$_1$, Q$_2$), showing near-degeneracy and the emergence of a conical intersection region. The arrow indicates a representative point in coordinate space. Representative normal modes corresponding to Q$_1$ and Q$_2$ are shown, illustrating the molecular distortions that drive the system toward the intersection region. (B) One-dimensional cuts of the potential energy surfaces along Q$_1$ and Q$_2$, respectively, highlighting the approach to degeneracy between the $^4$T$_2$ and $^2$E states along these nuclear coordinates. (C) Simulated population dynamics following excitation to the $^4$T$_2$ state, showing ultrafast population transfer to the $^2$E state. The dashed line indicates an exponential fit with a characteristic timescale of 153 fs.} 
\end{center}
\end{figure}

We carried out {\em ab-initio} calculations to elucidate the vibrational behavior and explore the excited-state dynamics of the molecule. Quantum chemical methods were applied to characterize both the electronic and vibrational features in the gas phase. All computations were performed using the ORCA \cite{orca} package, employing the B3LYP\cite{b3lyp}  functional in conjunction with the def2-TZVP\cite{def2tzvp} basis set. A comparison of geometries revealed that the relaxed structure with a spin multiplicity of 4 is energetically more favorable than that with a multiplicity of 2, thereby identifying it as the ground-state configuration in this study. The Raman spectrum was also computed using ORCA and is presented in Fig.\ S13 of the SI, alongside the experimentally measured spectrum. Vibrational modes contributing to prominent spectral features with discernible Raman activity are illustrated in Fig.\ S14, where atomic displacements are denoted by directional arrows. As shown in Fig.\ \ref{fig:Fig4}(a), two notable vibrational modes located at 256 cm$^{-1}$ and 451 cm$^{-1}$ correspond to O-Cr-O, Cr-O-C bending and Cr-O stretching coupled with C-CH-C bending, respectively. These modes exhibit significant vibronic coupling, quantified by fitting adiabatic potential energy surfaces (PESs) to the eigenstates of a vibronic Hamiltonian constructed in a diabatic representation. Details of the PES scans and the fitting methodology are provided in the Materials and Methods section and elaborated in Section III of the SI. The SOC strengths were also computed and benchmarked against prior results\cite{Hideo Ando 2012}, with all computed SOC values included in the SI.

Building upon the results of our \emph{ab initio} calculations, we developed a spin-vibronic coupling model incorporating refined parameters and a physically reasonable strength of SOC as elaborated in the Materials and Methods section and Section III of the SI. Key vibrational motions exhibiting relatively strong vibronic interactions were identified, and corresponding vibrational levels were selected for constructing the model Hamiltonian. The interplay of electronic, vibrational, and spin degrees of freedom leads to a high-dimensional system Hamiltonian. To investigate the population transfer dynamics in the [Cr(acac)$_3$] complex, we employed the hierarchical equations of motion (HEOM) formalism\cite{HEOM1, HEOM2}, which provides numerically exact solutions. Numerical calculations were carried out using an RTX4090 GPU to handle the intensive computational demands.  In alignment with experimental observations, two key vibrational modes: 256 and 451 cm$^{-1}$ were selected for modeling given their significant vibronic coupling strengths. Population dynamics simulations were conducted and are presented in Fig.\ \ref{fig:Fig4}(C), assuming an initial population localized in the lowest bright excited state, $^{4}$T$_{2}$. To evaluate the effects of mode combinations, we first examined the dynamics induced by pairing the 451 and 256 cm$^{-1}$ modes. The strong vibronic interactions and SOC between relevant excited states facilitated efficient excited state dynamics, with a characteristic timescale of 153 fs, obtained through exponential decay fitting. Similarly, simulations with the 451 and 231 cm$^{-1}$ mode combination yielded a transfer timescale of 131 fs, while the pairing of 451 and 681 cm$^{-1}$ resulted in a timescale of 124 fs (SI, Fig. S17). These findings indicate that the transition from the $^{4}T_{2}$ to the $^{2}$E occurs on a sub-200 fs timescale, consistent with results determined from 2DES. Additionally, when all three modes (231, 451, and 681 cm$^{-1}$) were included simultaneously, the calculated population dynamics revealed a slightly faster transfer with a lifetime of 120 fs, indicating a minor enhancement in transfer efficiency compared to two-mode scenarios. Notably, vibrational coherences were also observed at early waiting times, which decayed rapidly within 150 fs. To elucidate the complexity of the population transfer mechanism, we introduce a conceptual framework based on PESs. Specifically, we examine how two electronic states, coupled both vibronically through key vibrational modes (451 and 256 cm$^{-1}$) and electronically via SOC, give rise to avoided crossings in the nonadiabatic basis upon diagonalization of the system Hamiltonian. The resulting PESs, shown in Fig.\ \ref{fig:Fig4}(A), are plotted along two selected vibrational coordinates (denoted Q$_{1}$ and Q$_{2}$ in Fig.\ \ref{fig:Fig4}(B)). Applying the same methodology, we generated PESs for another vibrational mode pair (451 and 681 cm$^{-1}$) (SI, Fig. S22). Notably, these surfaces reveal the presence of strong vibronic interactions and SOC that facilitate the formation of an effective conical intersection between the electronic states.

To gain a deeper mechanistic understanding of the interplay between SOC and vibronic interactions in mediating electronic population transfer, we conducted a systematic exploration of how variations in coupling strengths influence the nonadiabatic dynamics within the [Cr(acac)$_{3}$] complex. Specifically, we modified the vibronic coupling parameters $\kappa$ and $\lambda$ for each selected mode. Here, $\kappa$ denotes the vibronic coupling strength associated with modulation of site energies, thereby influencing the shape and curvature of the PESs and affecting nonadiabatic interactions \cite{Tuning mode}. In contrast, $\lambda$ characterizes the vibronic coupling that modulates electronic couplings by facilitating direct mixing of electronic wave functions between states \cite{Coupling mode}. We also examined the effects of varying SOC strength. The results of these calculations are presented in Section III of the SI. Population dynamics simulations with scaled $\kappa$ values (0.8× and 1.2× of the refined reference value) for the 451 and 231 cm$^{-1}$ mode combination are displayed in Fig.\ S18. Compared with reference data from the main text, we observed that the transfer rate does not scale linearly or predictably with the absolute magnitude of $\kappa$, underscoring the mechanistic complexity of the transfer process. However, this nonlinearity can be qualitatively interpreted using the conical intersection framework: changes in $\kappa$ do not simply shift the degeneracy point between PESs but instead induce nonlinear effects. A parameter regime for $\kappa$ yielding optimized transfer speed was identified. In contrast, variations in $\lambda$ yielded more interpretable results. Increasing $\lambda$ led to a clear enhancement in transfer efficiency between the electronic states, consistent with its role in directly mixing electronic wavefunctions. Additionally, we explored the influence of SOC strength on transfer dynamics. The simulations revealed that increasing SOC generally decreases transfer efficiency within the [Cr(acac)$_{3}$] complex, as supported by the fitted timescales. 

Further investigations were carried out for additional vibrational mode combinations: (451, 256 cm$^{-1}$) and (451, 681 cm$^{-1}$), with population dynamics and corresponding fitting procedures presented in Figs.\ S19 and S20 of the SI. These results reinforce the notion that the relationship between the parameters $\kappa$, $\lambda$, and transfer rates is not trivially monotonic, owing to the intricate interplay of vibronic couplings and SOC. Given its relatively simple electronic structure and limited number of participating states, the [Cr(acac)$_{3}$] complex serves as an ideal model system for probing ISC mechanisms in a controlled setting. To further understand how the degeneracy point between PESs responds to variations in SOC and vibronic couplings, we systematically analyzed the energy changes of this point in the adiabatic representation following diagonalization of the system Hamiltonian. These results are provided in Figs.\ S21 and S22 of the SI.

\subsection{Discussion} 

By directly correlating vibrational coherences with evolving electronic spectral features, our work resolves a central gap highlighted in prior studies of  [Cr(acac)$_{3}$]\cite{Ref30}, where coherent motion was detected but not mechanistically linked to spin conversion pathways . Our combined 2DES-TG-time-frequency framework reveals that low-frequency metal-ligand modes (255 and 477 cm$^{-1}$) are not incidental, but actively track excited-state absorption and ground-state bleaching within the sub-picosecond ISC window. Unlike earlier assignments of isolated coherences (e.g., 164 cm$^{-1}$) without spectral or temporal mapping, we demonstrate that these modes persist and co-localize with population transfer, establishing their role as effective reaction coordinates embedded within the electronic manifold. This enables a unified mechanistic picture in which photoexcitation launches a vibronic wavepacket on the $^{4}$T$_{2}$ surface, and nuclear motion along specific metal-ligand coordinates simultaneously tunes energy gaps and enhances electronic mixing. More specifically, our spin-vibronic model suggests that the two dominant low-frequency modes play complementary roles in steering the wavepacket towards the intersection seam. The 255 cm$^{-1}$ O-Cr-O bending motion primarily acts as a tuning coordinate, periodically modulating the relative energies of the $^4$T$_2$ and $^2$E potential energy surfaces and reducing their energy separation. Simultaneously, the 477 cm$^{-1}$ Cr-O stretching mode functions predominantly as a coupling coordinate, perturbing the metal-ligand orbital overlap and increasing vibronic mixing between the electronic states. Their concerted motion therefore drives the nuclear wavepacket through a multidimensional region where the energy gap collapses while interstate coupling is maximised, generating an effective conical-intersection seam that provides an ultrafast pathway for spin exchange despite the intrinsically weak spin-orbit coupling of Cr(III). As the wavepacket evolves towards the multidimensional intersection seam, the tuning coordinate reduces the $^4$T$_2$–$^2$E energy gap while the coupling coordinate maximises interstate mixing. Under these conditions, nonadiabatic coupling reaches its maximum and even the intrinsically weak spin-orbit interaction becomes sufficient to efficiently admix quartet and doublet characters, enabling ultrafast $^4$T$_2$ $\rightarrow$ $^2$E transfer. Importantly, the sustained coherence of these low-frequency metal-ligand modes directly regulates both the rate and yield of spin conversion by maintaining the system within a vibrationally steered region of near-degeneracy where nonadiabatic coupling and SOC act cooperatively. This establishes a predictive framework for 'mode engineering,' wherein targeted modification of ligand rigidity, symmetry, and mass distribution can selectively tune these active coordinates. By explicitly connecting vibrational dynamics with PES topology and spin-state mixing, our results provide a mechanistic foundation for rationally controlling light-induced spin switching in earth-abundant paramagnetic 3d transition metal complexes, advancing their functional integration in molecular spintronic applications.


\section*{Conclusion}

Ultrafast molecular spin control ultimately requires linking a measured spin‑state conversion timescale to the nuclear coordinates that make the spin flip possible. Our work illustrates how combining frequency‑resolved ultrafast spectroscopy with ab initio-parameterized spin‑vibronic models can deliver that link: spectroscopy localizes where coherences and population dynamics reside spectrally, while modeling translates those observables into PES topology, coupling coordinates, and intersection accessibility. The implication is a practical design rule for earth‑abundant 3d metals: when SOC is weak, engineered vibronic coupling can compensate by steering wavepackets into near‑degeneracy and enhancing electronic mixing. Ligand modification and symmetry control can reshape metal-ligand modes and their $\kappa/\lambda$ character; rigidity tuning, bite‑angle changes, and targeted isotopic substitution provide additional handles to shift frequencies, displacements, and couplings. Such “mode engineering” offers a route to accelerate, slow, or spectrally gate spin conversion, with direct relevance to molecular spintronics (creating fast, switchable spin manifolds). More broadly, these results highlight PES topology as an actionable target alongside electronic structure in designing next‑generation photoactive complexes. Iterating this workflow across ligand frameworks can establish predictive structure-dynamics design rules.

%


\section*{Materials and Methods} 
\subsection{Sample preparation.} 

The $\rm [Cr(acac)_{3}$] complex was ordered from YuanYe Bio company (website: http://www.shyuanye.com). The complex was initially dissolved in methanol and subsequently diluted to achieve an optical density (OD) of 0.2 at 560 nm, corresponding to the absorption maximum. A micropump system (Mcropump GA-X21-DEMSE) was employed to circulate the sample through a flow cell, with a 1 mm pathlength, ensuring continuous sample renewal during measurements. A custom-built XYZ delay-stage system was implemented to precisely control the spatial alignment and focal position of the laser beam on the sample. This configuration allowed for fine-tuning of beam size and sample positioning, thereby optimizing the conditions for TA, TG, and 2DES measurements. Maximization of the signal-to-noise ratio was achieved by minimizing light scattering. 

\subsection{TG and 2DES measurements} 

Details of the experimental setup have been described in earlier reports from our group \cite{Ref2D1, Ref2D}. Briefly, the measurements were conducted using an all-reflective, diffractive optics-based 2D spectrometer, which provides exceptional phase stability of approximately $\lambda$/160. The excitation source was a home-built nonlinear optical parametric amplifier (NOPA), pumped by a commercial femtosecond laser system (Spectra Physics, Newport). The NOPA output was compressed to $\sim$10 fs pulse duration using a combination of a 19-channel deformable mirror (OKO Technologies) and a fused silica prism pair. Temporal characterization of the compressed pulses was performed via frequency-resolved optical gating (FROG), and the resulting traces were analyzed using the commercial software FROG3 (Femtosecond Technologies). The broadband laser spectrum exhibited a full-width at half-maximum (FWHM) of $\sim$120 nm, centered around 610 nm, effectively encompassing the transition to the first excited electronic state of the complex. For 2DES measurements, three collinear pulses were focused onto the sample with a spot size of $\sim$130 $\mu$m, and the resulting photon echo signal was collected in the phase-matching direction. Detection was carried out using a Sciencetech model 9055F spectrometer coupled to a linear CCD array camera (Entwicklungsb{\"u}ro Stresing). 

TG measurements were acquired by scanning the population (waiting) time delay over a range from –200 fs to 3000 fs, using a time step of 3 fs. To enhance the signal-to-noise ratio, 8000 spectra were averaged for TA and 200 for TG at each delay point. For 2DES, the coherence time $\tau$ was scanned within a symmetric window of –64 to +64 fs, and the waiting time T was varied from 0 to 2000 fs in 10 fs increments. During all measurements, the excitation pulse energy was attenuated to 30 nJ, and experiments were conducted at a repetition rate of 1 kHz. Phasing of the obtained TG and 2DES spectra was performed using an ``invariant theorem", which has been described in Ref.\ \citeonline{David2001}. At time zero (T = 0 fs), a strong negative signal was observed, which we attribute to non-resonant contributions arising from the overlap of pump and probe pulses, as well as potential substrate interactions. This artifact, while prominent, was short-lived and rapidly decayed within the instrument response window. To ensure the integrity of the dynamic signal, we excluded data from T = 0 to T = 40 fs in our subsequent analysis and initiated our evaluation from T = 40 fs onward, where pulse overlap effects no longer influence the measurements. 

\subsection{Theoretical calculations.}  

The optimized molecular structures, the normal modes, and the Raman spectrum of the molecule are obtained via density functional theory calculations on the level of theory B3LYP/def2-TZVP as implemented in the ORCA package\cite{orca}. For the PES scan along a vibraitonal mode, the energies of the $^4T_2$ and $^2E$ states are obtained using the same functional and basis set, with the details and results presented in the SI. 

Based on the {\em ab-initio} calculations, we were able to construct a spin-vibronic-coupling model for the Cr complex. The system Hamiltonian can be written as 
\begin{equation}
\begin{split}
\label{eq:system Ham} 
H_{s} =& \sum^{N}_{i=1}\ket{e_{i}}(\epsilon_{i}+h_{i})\bra{e_{i}} + \sum_{i\neq j}\ket{e_{i}}(V_{ij}+\lambda_{ij}Q_{c})\bra{e_{j}}, \\
h_{i} =& \frac{1}{2}\hbar\Omega_{t}(\alpha^{\dagger}_{t}\alpha_{t}+\frac{1}{2}) +\frac{1}{2}\hbar\Omega_{c}(\alpha^{\dagger}_{c}\alpha_{c}+\frac{1}{2}) + \kappa_{i}Q_{t}, 
\end{split}
\end{equation} 
where, $\ket{e_{i}}$ is the electronic singlet or triplet states. Here, it denotes the S$_{1}$ and S$_{2}$ ($^4$T$_{2}$ and $^2$E), respectively. Thus, N = 2 in this work. V$_{ij}$ and $\lambda_{ij}$ are the parameters of the SOC and the vibronic coupling of the coupling mode, which tune the couplings between two electronic states. Moreover, $\kappa_{i}$ is the parameter determining vibronic coupling of the tunning mode, which tune the site energies of electronic states. It shows the relation $\kappa_{i}$ = $\Delta_{i}\Omega_{t}/\sqrt{2}$, where $\Delta_{i}$ is the dimensionless shift of the minimum position of the i-th potential energy surface comparing to the position of ground electronic state. $\Omega_{t}$ and $\Omega_{c}$ are the frequencies of the tuning and coupling mode, respectively. In this work, we employed the system-bath model to consider the dissipation and relation acted by noisy environment. Thus, we have $H = H_{s}+H_{env}$ and the environment can be considered as an infinity number of harmonic oscillators, which acts as a thermal reservoir. The environmental bath Hamiltonian is given as 
\begin{eqnarray}
\label{eq:molecular Hamiltonian}
H_{\rm env} &=& \sum_{\alpha}\Big[\frac{p^{2}_{\alpha}}{2m_{\alpha}} +\frac{m_{\alpha}\omega^{2}_{\alpha}}{2} \left( 
x_{\alpha}+\frac{c_{\alpha}\ket{S_1}\bra{S_1}}{m_{\alpha}\omega^{2}_{\alpha}}  \right)^{2} \nonumber \\ 
& & +   \frac{q^{2}_{\alpha}}{2M_{\alpha}}+\frac{M_{\alpha}\nu^{2}_{\alpha}}{2}
\left( y_{\alpha}+\frac{t_{\alpha}\ket{S_{2}}\bra{S_{2}} }{M_{\alpha}\nu^{2}_{\alpha}} \right)^{2}\Big]\, .
\end{eqnarray}
Here, the momenta of the bath oscillators are denoted as $p_{\alpha}$ and $q_{\alpha}$, while their coordinates, masses, and frequencies are denoted by $x_{\alpha}, m_\alpha, \omega_\alpha$, and   $y_{\alpha}, M_\alpha, \nu_\alpha$. The respective coupling constants are $c_\alpha$ and $t_\alpha$. The baths are characterized by the spectral densities $J_{\rm c}(\omega)=\frac{\pi}{2}\sum_\alpha\frac{c_\alpha^2}{m_\alpha\omega_\alpha} \delta(\omega-\omega_\alpha)$ and $J_{\rm t}(\omega)=\frac{\pi}{2}\sum_\alpha\frac{t_\alpha^2}{M_\alpha\nu_\alpha} \delta(\omega-\nu_\alpha)$. Throughout this work, we assume two Loretizian spectral densities, which show the form $J(\omega) = 2\lambda'\frac{\omega\gamma}{\omega^{2}+\gamma^{2}}$. The reorganization energy $\lambda' = \frac{\zeta\gamma}{2\pi}$ and cutoff frequency $\gamma$. The model parameters were obtained based on the {\em ab-initio} calculations, the detailed values of parameters are described in the SI.


%
\begin{addendum} 
\item This work was supported by the National Key Research and Development Program of China (Grant No.\ 2024YFA1409800), NSFC Grants with No.\ 12274247 and 12504279, Zhejiang Provincial Natural Science Foundation of China with No. LQN26A040009, Yongjiang talents program with No.\ 2022A-094-G and 2023A-158-G, Ningbo International Science and Technology Cooperation with No.\ 2023H009, `Lixue+' Innovation Leading Project and the foundation of national excellent young scientist.

\item[Supporting information] It provides comprehensive details supporting the experimental and theoretical results. It includes transient absorption measurements and advanced data analysis methods such as global fitting of 2DES datasets, extraction of decay-associated spectra, and construction of two-dimensional vibrational maps. Time-domain residual analysis, Fourier transforms, and wavelet methods are used to resolve coherent vibrational dynamics. Theoretical sections present quantum chemistry calculations (DFT, TDDFT, and CASSCF), Raman spectra comparisons, and potential energy surface scans. Additionally, it outlines the spin-vibronic coupling model, parameter refinement, and quantum dynamics simulations, highlighting the roles of vibronic interactions and spin–orbit coupling in intersystem crossing. 

\item[Competing Interests] The authors declare that they have no competing financial interests. 

\item[Correspondence] Correspondence of paper should be addressed to F. Z. ~(email:zhengfulu@nbu.edu.cn), A. J. (Ajay.Jha@rfi.ac.uk) and H.-G.D. ~(email: duanhongguang@nbu.edu.cn). 

\end{addendum}
%
\newpage

\newpage

\newpage

\newpage

\end{document}